\documentclass[conference]{IEEEtran}
\IEEEoverridecommandlockouts
\usepackage{cite}
\usepackage{amsmath,amssymb,amsfonts}
\usepackage{algorithmic}
\usepackage{graphicx}
\usepackage{textcomp}
\usepackage{hyperref}
\usepackage{listings}
\usepackage{tikz}
\def\checkmark{\tikz\fill[scale=0.2](0,.35) -- (.25,0) -- (1,.7) -- (.25,.15) -- cycle;} 
\usepackage{hyperref}
\usepackage{xcolor}
\def\BibTeX{{\rm B\kern-.05em{\sc i\kern-.025em b}\kern-.08em
    T\kern-.1667em\lower.7ex\hbox{E}\kern-.125emX}}
\begin{document}

\title{DoReMi: First glance at a universal OMR dataset \\
\thanks{Funded by UKRI, EPSRC and Steinberg.}
}

\author{\IEEEauthorblockN{1\textsuperscript{st}Elona Shatri}
\IEEEauthorblockA{\textit{Centre for Digital Music} \\
\textit{Queen Mary University of London}\\
London, UK \\
e.shatri@qmul.ac.uk}
\and
\IEEEauthorblockN{2\textsuperscript{nd} Gy\"orgy Fazekas}
\IEEEauthorblockA{\textit{Centre for Digital Music} \\
\textit{Queen Mary University of London}\\
London, UK \\
g.fazekas@qmul.ac.uk}
}

\maketitle

\begin{abstract}
The main challenges of Optical Music Recognition (OMR) come from the nature of written music, its complexity and the difficulty of finding an appropriate data representation. This paper provides a first look at DoReMi, an OMR dataset that addresses these challenges, and a baseline object detection model to assess its utility. Researchers often approach OMR following a set of small stages, given that existing data often do not satisfy broader research. We examine the possibility of changing this tendency by presenting more metadata. Our approach complements existing research; hence DoReMi allows harmonisation with two existing datasets, DeepScores and MUSCIMA++.  DoReMi was generated using a music notation software and includes over 6400 printed sheet music images with accompanying metadata useful in OMR research. Our dataset provides OMR metadata, MIDI, MEI, MusicXML and PNG files, each aiding a different stage of OMR. We obtain 64\% mean average precision (mAP) in object detection using half of the data. Further work includes re-iterating through the creation process to satisfy custom OMR models. While we do not assume to have solved the main challenges in OMR, this dataset opens a new course of discussions that would ultimately aid that goal.

\end{abstract}

\begin{IEEEkeywords}
optical music recognition, deep learning, dataset, sheet music
\end{IEEEkeywords}

\section{Introduction}

Despite improvements in music notation software, writing music on paper or distributing music in print is still very common. However, written music needs to be digitised for further editing, preparation for printing, sharing or auditioning on a computer, or creating playable musical demos. The process of manually digitising scores is slow and tiresome. New work is not the only subject of this process. The abundance of undigitised sheet music in archives and libraries is an essential motivation of OMR as well. Scores are often scanned and become part of library archives which advances their accessibility only partially. For example, the search of scanned music is limited to metadata and excludes musical content or patterns.
Moreover, scanned scores do not allow for plagiarism check. An automatic process of converting handwritten scores, prints and scans to documents that a machine can read and interpret is the essence of Optical Music Recognition (OMR). OMR research typically divides the problem into distinct stages. Conventionally, four main stages are considered, starting with image pre-processing, followed by the detection of primitive musical objects, reconstructing musically meaningful composites, and finally, encoding using a machine-readable format \cite{b24}. By progressing one stage at a time, inconsistencies were created, mainly in the datasets used by researchers. Often, new work does not align with current work creating issues in evaluation and comparison. 

Recently, researchers have been working on bringing their datasets together, namely MUSCIMA++~\cite{b4} and DeepScores~\cite{b9}, to improve compatibility. MUSCIMA++ contains handwritten music, while DeepScores is a typeset music dataset, both of which can be used in object detection. Harmonising them aids the goal of creating a reference dataset. With the motivation to support this goal and compare and use these datasets in future experiments, DoReMi is easily harmonised with MUSCIMA++ and DeepScores. 
Another limitation of several existing OMR datasets is that only one research stage is supported primarily, while other stages, such as reconstruction, are typically understudied. DoReMi addresses this issue by also providing semantic information of the written music. Given that DoReMi was generated using music notation software, we could retrieve musical information that otherwise would be impossible to obtain. Such data is available for playable notes (also grace notes), stave lines, clefs, slurs and ties. Depending on the nature of the element, different degrees of semantic information is provided. This data is integrated with visual data, such as bounding boxes and pixel information of each element. We expect this to facilitate further research in end-to-end deep learning systems in OMR, which are quickly becoming state-of-the-art. 
Furthermore, DoReMi includes different types of representations, including MusicXML, MIDI, MEI and PNG images of sheet music alongside OMR metadata.

\section{Handwritten and Typeset OMR Datasets}

Most of the existing OMR datasets target different approaches, stages and objectives within OMR research. Differences between datasets primarily lie in the annotations that accompany images of sheet music. There are also variations in the type of notation, such as CWMN or mensural notation, in primary focus, and differences in the printing type, i.e., typeset or handwritten scores. A list of the main datasets, engraving type, number of symbols and A4 images, formats provided and usage is given in Table \ref{table:OMR-datasets}.
Conditional on the final objective of the task, there are also different music representations used in datasets. For instance, we only need the MIDI file as the ground truth to retrieve a replayable-only file. If the graphical elements and an editable score are needed, other more complex formats such as MusicXML and MEI are essential. The use case determines the optimal music representation. MEI and MusicXML are score focused, while MIDI focuses on the musical content and synthesiser control.


HOMUS (Handwritten Online Musical Symbols) \cite{b6} takes the approach of recording pen-based (online) compositions. They present the strokes drawn by pen and the image generated after drawing the symbols. The authors then propose a baseline on the recognition of these two modalities. To conclude the recognition accuracy, they use Nearest Neighbor (NN) technique and Hidden Markov Models (HMM).

CVC-MUSCIMA \cite{b8} is the root dataset of MUSCIMA++. CVC-MUSCIMA was originally designed to perform stave line removal. A total of 50 different musicians were asked to write 20 pages of identical sheet music, having the same pen and style. This dataset creates a link between low-level such as noteheads and higher-level symbols such as key and time signatures and is mainly used and best suited for musical object detection. Baseline experiments for object detection using MUSCIMA++ use detection algorithms such as Faster R-CNN \cite{b3}, Single Shot Detectors \cite{b22} and DeepWaterShed Detectors \cite{b23}. A version of MUSCIMA++ containing annotations for measures and staves only exists for bar measure detection. 

Another dataset that assists in the detection stage is DeepScores \cite{b9} which is the largest OMR dataset that contains annotated images of typeset scores used for object classification, detection and segmentation. There is variation provided by rendering the sheets using five different fonts. This dataset assists work in recognising tiny objects in large images. A new version of DeepScores has detailed annotations, increased number of annotated symbols, while also providing oriented bounding boxes for the symbols, a higher level of rhythm and pitch information that includes onset beat for all symbols and line position for noteheads and finally compatibility with MUSCIMA++ dataset~\cite{b19}.

Printed Images of Music Staves (PrIMuS)~\cite{b10} is one of the few datasets that satisfy use cases in training end-to-end object recognition models. It includes 87,678 real-music sequences of notes, typically the first ones, in five different formats: MIDI, PNG, MEI, semantic and an encoding that contains the symbols and their positions, disregarding their musical meaning. Another version Camera-PrIMuS~\cite{b11} includes images with distortion to simulate real-world imperfections. 

Universal Music Symbol Collection~\cite{b7} is a dataset that collects and combines symbols from HOMUS, MUSCIMA++, Audiveris OMR dataset, the Printed Music Symbols dataset, OpemOMR dataset and two sets from the group of Rebelo et al. \cite{b15} and Fornes et al. \cite{b16} that can be used to train classifiers. The symbols amount to 74,000 handwritten and 16,000 printed symbols. The objective was to create a universal, harmonised dataset that could assist in building a written music classifiers. 

Limitations of each dataset lie in their differences in data types, formats, and narrow objectives. DoReMi goes one step further by allowing harmonisation with MUSCIMA++ and DeepScores while adding semantic and graphical information about the symbols. Furthermore, DoReMi uses five file formats, XML (with positions), MusicXML, PNG, MEI and MIDI, with complementary information. PNG images are complemented by metadata in an XML metadata file. MEI, MusicXML and MIDI all representing possible encoded outputs of OMR, depending on the task.


\begin{table*}[!htbp]
    \centering
    \caption{Comparison of major OMR datasets published to date}
    \begin{tabular}{|p{3.1cm}|p{1.5cm}|p{1cm}|p{0.8cm} |p{0.8cm}|p{3.3cm}|p{4cm}|}
    \hline \hline
    Dataset  & Engraving  & Symbols & Images & Classes & Format  & Usage  \\
    \hline 
    DoReMi & Typeset & 911771 & 6432 & 94 & \checkmark ML metadata, images, MIDI, MEI, MusicXML & 
    Object Detection, Reconstruction and Encoding, End-to-end  \\ 
    \hline
    Handwritten Online Musical Symbols (HOMUS) \cite{b6} & Handwritten & 15200 & - & 32 & Text-File & 
    Symbol Classification (online + offline) \\ 
    \hline 
    Universal Music Symbol Collection \cite{b7}  & Typeset + Handwritten & 90000 & - & 79 & Images  & Symbol Classification (offline) \\ 
    \hline 
    MUSCIMA ++ \cite{b4}   & Handwritten &  91255 & 140 & 110  & Images, Measure Annotations, MuNG   & Symbol Classification, Object Detection and Measure Recognition \\ 
    \hline
    DeepScores \cite{b9, b19}  & Typeset  &  100m & 255,386  & 135 & Images, XML  & Symbol Classification, Object Detection, Semantic Segmentation  \\ 
    \hline
    PrIMuS \cite{b11} &  Typeset & 87678 & - & - & Images, MEI, simplified encoding, agnostic encoding & End-to-End Recognition        \\ 
    \hline
    Capitan collection \cite{b20} & 
    Handwritten & - &
    10230 & 30  & Images, Text-File & Symbol Classification    \\ 
    \hline
    Bounding Box Annotations of Musical Measures \cite{b21}  & Typeset  & 
    940 & 24,329 & - &
    Images & 
    CSV, plain JSON and COCO  \\
    \hline
    \hline
    \end{tabular}
    \footnotesize
    \label{table:OMR-datasets}
\end{table*}

\section{Dataset Description}
In this section, we present DoReMi, an OMR dataset with typeset symbols. This dataset is designed to be compatible with MUSCIMA++ and DeepScores and serve as a reference for research in other stages of OMR. Notably, it helps research in the reconstruction stage of music semantics such as notes pitch, duration, beats and their relations to other elements in the score. Furthermore, we expect DoReMi to aid work towards an end-to-end OMR system in non-monophonic scores.

Music used to generate this dataset comes from a software test set provided by the Dorico team \footnote{Dorico is a music notation software - https://new.steinberg.net/dorico}. This test set includes a wider number of objects, classes and various cases of notations not normally seen in real-world music. About 600 files of the underlying material used to generate this dataset is copyright protected; therefore, we only include openly distributable scores in the final published dataset \footnote{\url{https://github.com/steinbergmedia/DoReMi/releases}}. However, pre-trained models trained in the whole dataset will also be published. DoReMi includes around 6432 images of sheet music with nearly a million annotated objects which is \(\frac{1}{50}\)th the size of DeepScores and 42 times the size of MUSCIMA++. Each object on the page is annotated with category labels from 94 different classes. However, there is an emphasised class imbalance; stems and noteheads make up half of the annotated objects in the dataset. We also provide prepared subsets fulfilling different requirements on the number of pages, number of classes and the number of staves. Most of the images include one system per page; depending on the number of voices, they have one or more staves per page.

Following the organisation of MUSCIMA++, DoReMi has an OMR metadata file which includes bounding boxes of each element: top, left, width and height. It also includes the pixel mask for each element giving each object's pixels inside the bounding box. Additionally, DoReMi provides the relationships between primitives. It vaguely follows the Music Notation Graph (MUNG), which creates a graph representation of music notations. Inlinks and outlinks reference back and forth to the ID of the objects they are related to—for instance, a notehead half outlinks to a stem or a slur or both of them. Conversely, the stem inlinks to the notehead half. 

As opposed to other existing datasets, DoReMi provides semantic information on playable notes and interpretive elements, see Appendix \ref{appendix}. Certain (playable) objects are also annotated with a Dorico event ID which is a unique event identifier that provides additional information on how some objects are linked. For instance, notes like noteheads have information on the duration beats, onset beats, pitch octave, midi pitch code, normalised pitch step and an event ID. For elements such as clefs, our dataset provides an event ID, clef type, clef hotspot, clef required stave lines and clef stave position. Clef hotspot identifies the midi pitch that clef denotes, i.e. for treble clef is G4, as that is the pitch of the second stave line from the bottom. Clef required stave lines shows how many stave lines the clef needs. Time signatures include the event ID and its description, for example, 3/2 (h, 1+1+1) \footnote{Beat division: h means half note, 1 + 1 + 1 means 3 equal beat, 1,2,3}. Flags, if they are part of grace notes, have a boolean value set to True. Slurs and ties have their event IDs, while barlines, rests, accidentals, augmentation dots, stems do not have such information. Beams do not have their event ID; instead, they have a list of the event IDs their respective noteheads have. 
Other types of data given are Dorico project files, MIDI files, PNGs, MusicXML and MEI. PNG files are binarised and provided with a resolution of 300 DPI and dimensions of 2475x3504 pixels. There is a possibility of creating lower or higher resolution images depending on limitations in computational expense. One OMR XML metadata file may be pointing back to multiple images. Each image has a reference page ID in the XML file.
MIDI files included can be used as ground truth to OMR tasks where MIDI is the desired output. MusicXML \cite{b17} and MEI \cite{b14} are two file formats that we desire to output after encoding the reconstructed information. They were conceived for two different reasons. MusicXML was first proposed as a file format to ease digital sheet exchange in the music publishing industry \cite{b18}. MEI was born in the music research world to aid the storage of diverse music manuscripts. Both share similarities in the objects they encode and their file format being XML. MEI, beyond functionality in notation and page layout, also encodes information about the notation in a more structured and semantic way. In other words, MusicXML was designed for software rendering, while MEI captures more music semantics \cite{b24}. 

\section{Baseline Experiments}\label{exp}

\begin{table*}[htbp]
\caption{Object detection baseline results in DoReMi and MUSCIMA++}
\begin{center}
\begin{tabular}{|llllll|}
\hline 
Meta-Architecture  & Feature Extractor  & Classes & Training steps & Data\% & mAP (\%)   \\  \hline \hline
\newline
DoReMi & & & & & \\
\hline
Faster R-CNN & Inception-ResNet-v2  & 71 & 80K & 90\%  & 57.5067 \\ \hline
Faster R-CNN &  Inception-ResNet-v2 with MUSCIMA++ & 71 & 120K &  90\% & 64.8614 \\ \hline
Faster R-CNN & ResNet50 & 71 & 120K  & 90\% & 63.4910 \\ \hline
Faster R-CNN & ResNet101  & 71 & 148K & 90\% &  26.994 \\ \hline \hline
\newline
MUSCIMA++ & & & & & \\ \hline
Faster R-CNN & Inception-ResNet-v2  & 110 & 80K & 100\%  & 82.4 \\ \hline
\end{tabular}
\footnotesize
\label{table:baseline-results}
\end{center}
\end{table*}

Object detection is a crucial stage of optical music recognition. It is concerned with localising and identifying objects of certain classes in a sheet music image. Using the DoReMi dataset, we propose a baseline in object detection to assess the benefits brought forward by the richer data in this task. Based on previous work from Pacha et al. \cite{b2, b6, b9}, we use Faster R-CNNs as our central architecture. We also compare the results with those reproduced using MUSCIMA++ using the same architecture. 

Detecting objects in sheet music is considered more challenging than many general-purpose computer vision tasks, given that the number of tiny objects is very high. To detect such objects, we need to localise objects that assist classification. This localisation is defined by four values $b_{x}, b_{y}, b_{h}$ and $b_{w}$, while the first two terms determine the centre of the bounding box that isolates that object, the latter ones provide its height and width. The network also provides a class name for classification purposes. When objects to be detected are not overlapping, it is reasonably easy to obtain and interpret their bounding boxes. When two objects overlap in the same grid cell, having more than one midpoint in a cell, the network needs to know which object to predict. This overlap of objects is often seen in sheet music. To deal with overlaps, anchor boxes can be used. Objects are assigned to the respective grid cell and an anchor box for that grid cell with the highest Intersection over Union (IoU). We get two predicted bounding boxes for each of the grid cells then omit low probability object predictions. Subsequently, for each class in our dataset, we run non-max suppression to generate predictions. Non-max suppression eliminates the bounding boxes that have a low probability, retaining the bounding box with the highest probability. 


Fast R-CNN \cite{b3} serves as a base for almost all proceeding work in object detection. Fast R-CNNs use selective search to generate region proposals, which is expensive. After region CNNs were introduced, Faster R-CNN became the state-of-the-art approach in object detection~\cite{b3}. Most of the work subsequent to it follows a similar architecture by adding other valuable blocks such as Mask R-CNNs, which also output the object masks indicating the pixels where the object is in the bounding box. Faster R-CNNs are very similar to Fast R-CNN, with the most significant difference being how the region proposals are considered and CNNs are run on those, which cut the run time. Faster R-CNNs introduce region proposal network (RPN), which enables sharing full-image convolution features with the detection network~\cite{b2}. RPN is a fully convolutional layer, trained to generate good region proposals used by the rest of the network.


The experiments \ref{table:baseline-results} are performed in a subset of the DoReMi dataset consisting of 5832 A4 images of scores. These images are of the same dimension throughout the dataset with the limitation of one system per page. These systems very often consist of multiple staves. The maximum number of staves per page is six. Moreover, files that had less frequent objects were disregarded for this set of experiment.

\subsection{Feature Extractors}

We first apply a convolutional feature extractor in all input images, so the highest-level features are retrieved. Based on the number of parameters and layer types, the processing time is highly affected. We show three types of feature extractors and their training time for reference. We use open source Inception Resnets (v2) \cite{b13} and Resnet50 and Resnet101 \cite{b12}. All use Tensorflow implementations. 
Inception Resnets (v2) is a blended extractor with 164 layers which benefits from optimisations of the residual connections and efficiency of Inception units. It replaces the filter concatenation stage of the Inception architecture with residual connections. Resnet50 is another feature extractor with a depth of 50 layers, meaning that it is considerably lighter than Inception Resnet (v2), hence faster. Resnet101 belongs to the same residual connection family with Resnet50 with more depth.

\subsection{Results}

A common technique to boost detection results is to feed the data into a pre-trained network using the image representations in intermediate layers. We use the respective pre-trained models in the general COCO dataset for each extractor, except for a Faster R-CNN meta-architecture that uses Inception Resnet (v2) with a pre-trained model on MUSCIMA++. This model provides the best mAP score, which can be a result of the same domain pre-training. Resnet50 performs very well, yielding a mAP of 63\% as shown in Table from where we can also see that Resnet50 produce better results. Furthermore training time using Resnet50 extractors is half of that using Inception Resnet (v2) \ref{table:baseline-results}.

\section{Discussions}\label{discussion}

Baseline work in object detection with deep learning uses pre-trained models such as Faster R-CNN, Fast R-CNN, SSD \cite{b1}, with some fine-tuning during training. However, while such pre-trained models can detect musical objects, classes that are not well-represented pose a significant challenge. Creating custom models for OMR is one of the main challenges for both the detection and note assembly stage, given the graphical and structured nature of music. Such models would improve the state-of-the-art work in OMR, but it would further facilitate and trigger new research paths.
The DoReMi dataset is designed in a way that it can be harmonised with Deepscore and MUSCIMA++, which allows various engravings and takes steps towards creating standardised, universal OMR datasets. The design can further help to standardise evaluation at different stages of OMR and facilitate end-to-end approaches in the field. The dataset design follows the mung (Music Notation Graph), where nodes represent primitives such as noteheads, stems, beams, and their relations being stored.
The DoReMi dataset is the first step towards overcoming some of the challenges mentioned above. DoReMi in baseline experiments does not perform as good as MUSCIMA++, resulting from MUSCIMA++ images being cut by staff, meaning the number of objects is often smaller and of a larger dimension. 
While this dataset aims to fill in existing gaps in OMR research, it is also prepared with a view to investigate a more extensive research question, i.e., whether deep learning can assist the OMR research field by providing data in richer structure, so common tasks considered in isolation may be gradually integrated into more complex architectures. What sets DoReMi apart from other existing datasets primarily is data richness. Given that this dataset was generated using Dorico, we could obtain more musical information and different data types. Finally, the pipeline developed for assembling this dataset allows us to extend the dataset, as well as iterate and re-design the structure if needed based on the course of new research.

\clearpage
\onecolumn
\appendices

\section{Dataset Samples and Individual Class Attributes}\label{appendix}

\begin{figure}[hbt!!]
\begin{center}
\centerline{
\includegraphics[width=17cm]{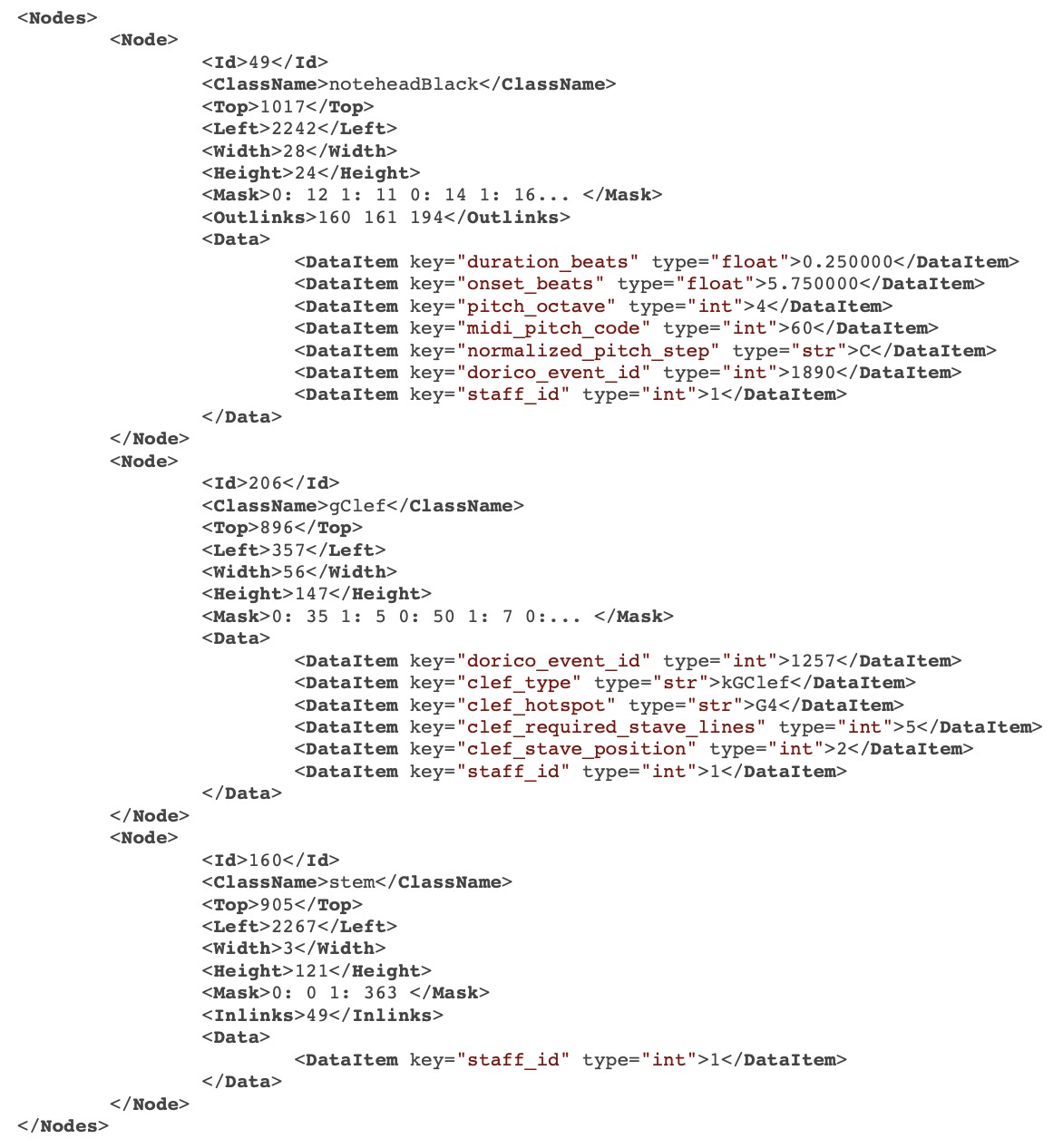}}
\caption{Snippet of three different nodes in OMR XML data files}
\label{fig:sample}
\end{center}
\end{figure}

\begin{table*}[h]
\caption{List of 94 classes showing frequency of appearance for each class and their data attributes}
\tiny
\begin{tabular}{|l|l|l|l|l|l|l|l|l|l|l|l|l|l|l|l|l|}
\hline
Class &
  Freq &
  DE id &
  clef type &
  clef hs &
  clef RSL &
  clef SP &
  staff id &
  DE ids &
  grace note &
  DB &
  OB &
  PO &
  MPC &
  NPS &
  TSD &
  text \\
 \hline
accidentalDoubleFlat                      & 244    &   &   &   &   &   & \checkmark &   &   &   &   &   &   &   &   &  \\
\hline
accidentalDoubleSharp                     & 330    &   &   &   &   &   & \checkmark &   &   &   &   &   &   &   &   &  \\
\hline
accidentalFlat                            & 12705  &   &   &   &   &   & \checkmark &   &   &   &   &   &   &   &   &  \\
\hline
accidentalKomaFlat                        & 5      &   &   &   &   &   & \checkmark &   &   &   &   &   &   &   &   &  \\ \hline
accidentalKomaSharp                       & 5      &   &   &   &   &   & \checkmark &   &   &   &   &   &   &   &   &  \\ \hline
accidentalNatural                         & 11137  &   &   &   &   &   & \checkmark &   &   &   &   &   &   &   &   &  \\ \hline
accidentalQuarterToneFlatStein            & 162    &   &   &   &   &   & \checkmark &   &   &   &   &   &   &   &   &  \\ \hline
accidentalQuarterToneSharpStein           & 191    &   &   &   &   &   & \checkmark &   &   &   &   &   &   &   &   &  \\ \hline
accidentalSharp                           & 12908  &   &   &   &   &   & \checkmark &   &   &   &   &   &   &   &   &  \\ \hline
accidentalThreeQuarterTonesFlatZimmermann & 1      &   &   &   &   &   & \checkmark &   &   &   &   &   &   &   &   &  \\ \hline
accidentalThreeQuarterTonesSharpStein     & 27     &   &   &   &   &   & \checkmark &   &   &   &   &   &   &   &   &  \\ \hline
accidentalTripleFlat                      & 1      &   &   &   &   &   & \checkmark &   &   &   &   &   &   &   &   &  \\ \hline
accidentalTripleSharp                     & 1      &   &   &   &   &   & \checkmark &   &   &   &   &   &   &   &   &  \\ \hline
articAccentAbove                          & 1477   &   &   &   &   &   & \checkmark &   &   &   &   &   &   &   &   &  \\ \hline
articAccentBelow                          & 1916   &   &   &   &   &   & \checkmark &   &   &   &   &   &   &   &   &  \\ \hline
articMarcatoAbove                         & 278    &   &   &   &   &   & \checkmark &   &   &   &   &   &   &   &   &  \\ \hline
articMarcatoBelow                         & 34     &   &   &   &   &   & \checkmark &   &   &   &   &   &   &   &   &  \\ \hline
articStaccatissimoAbove                   & 536    &   &   &   &   &   & \checkmark &   &   &   &   &   &   &   &   &  \\ \hline
articStaccatissimoBelow                   & 414    &   &   &   &   &   & \checkmark &   &   &   &   &   &   &   &   &  \\ \hline
articStaccatoAbove                        & 5018   &   &   &   &   &   & \checkmark &   &   &   &   &   &   &   &   &  \\ \hline
articStaccatoBelow                        & 6108   &   &   &   &   &   & \checkmark &   &   &   &   &   &   &   &   &  \\ \hline
articTenutoAbove                          & 823    &   &   &   &   &   & \checkmark &   &   &   &   &   &   &   &   &  \\ \hline
articTenutoBelow                          & 735    &   &   &   &   &   & \checkmark &   &   &   &   &   &   &   &   &  \\ \hline
augmentationDot                           & 2762   &   &   &   &   &   &   &   &   &   &   &   &   &   &   &  \\ \hline
barline                                   & 28142  &   &   &   &   &   &   &   &   &   &   &   &   &   &   &  \\ \hline
beam                                      & 52539  &   &   &   &   &   & \checkmark & \checkmark &   &   &   &   &   &   &   &  \\ \hline
cClef                                     & 1161   & \checkmark & \checkmark & \checkmark & \checkmark & \checkmark & \checkmark &   &   &   &   &   &   &   &   &  \\ \hline
dynamicFF                                 & 242    &   &   &   &   &   & \checkmark &   &   &   &   &   &   &   &   &  \\ \hline
dynamicFFF                                & 88     &   &   &   &   &   & \checkmark &   &   &   &   &   &   &   &   &  \\ \hline
dynamicFFFF                               & 4      &   &   &   &   &   & \checkmark &   &   &   &   &   &   &   &   &  \\ \hline
dynamicForte                              & 777    &   &   &   &   &   & \checkmark &   &   &   &   &   &   &   &   &  \\ \hline
dynamicFortePiano                         & 60     &   &   &   &   &   & \checkmark &   &   &   &   &   &   &   &   &  \\ \hline
dynamicForzando                           & 35     &   &   &   &   &   & \checkmark &   &   &   &   &   &   &   &   &  \\ \hline
dynamicMF                                 & 1803   &   &   &   &   &   & \checkmark &   &   &   &   &   &   &   &   &  \\ \hline
dynamicMP                                 & 185    &   &   &   &   &   & \checkmark &   &   &   &   &   &   &   &   &  \\ \hline
dynamicPiano                              & 3987   &   &   &   &   &   & \checkmark &   &   &   &   &   &   &   &   &  \\ \hline
dynamicPP                                 & 1807   &   &   &   &   &   & \checkmark &   &   &   &   &   &   &   &   &  \\ \hline
dynamicPPP                                & 93     &   &   &   &   &   & \checkmark &   &   &   &   &   &   &   &   &  \\ \hline
dynamicPPPP                               & 13     &   &   &   &   &   & \checkmark &   &    &   &   &   &   &   &   &  \\ \hline
dynamicRinforzando2                       & 5      &   &   &   &   &   & \checkmark &   &   &   &   &   &   &   &   &  \\ \hline
dynamicSforzato                           & 236    &   &   &   &   &   & \checkmark &   &   &   &   &   &   &   &   &  \\ \hline
dynamicSforzatoFF                         & 6      &   &   &   &   &   & \checkmark &   &   &   &   &   &   &   &   &  \\ \hline
dynamicText                               & 110    &   &   &   &   &   & \checkmark &   &   &   &   &   &   &   &   & \checkmark  \\ \hline
fClef                                     & 3543   & \checkmark & \checkmark & \checkmark & \checkmark & \checkmark & \checkmark &   &   &   &   &   &   &   &   &  \\ \hline
flag16thDown                              & 430    &   &   &   &   &   & \checkmark &   & \checkmark &   &   &   &   &   &   &  \\ \hline
flag16thUp                                & 18245  &   &   &   &   &   & \checkmark &   & \checkmark &   &   &   &   &   &   &  \\ \hline
flag32ndDown                              & 31     &   &   &   &   &   & \checkmark &   & \checkmark &   &   &   &   &   &   &  \\ \hline
flag32ndUp                                & 7264   &   &   &   &   &   & \checkmark &   & \checkmark &   &   &   &   &   &   &  \\ \hline
flag64thUp                                & 10     &   &   &   &   &   & \checkmark &   & \checkmark &   &   &   &   &   &   &  \\ \hline
flag8thDown                               & 3672   &   &   &   &   &   & \checkmark &   & \checkmark &   &   &   &   &   &   &  \\ \hline
flag8thUp                                 & 8113   &   &   &   &   &   & \checkmark &   & \checkmark &   &   &   &   &   &   &  \\ \hline
gClef                                     & 12278  & \checkmark & \checkmark & \checkmark & \checkmark & \checkmark & \checkmark &   &   &   &   &   &   &   &   &  \\ \hline
gradualDynamic                            & 7084   &   &   &   &   &   & \checkmark &   &   &   &   &   &   &   &   &  \\ \hline
kStaffLine                                & 160426 &   &   &   &   &   & \checkmark &   &   &   &   &   &   &   &   &  \\ \hline
mensuralNoteheadMinimaWhite               & 9      & \checkmark &   &   &   &   & \checkmark &   &   & \checkmark & \checkmark & \checkmark & \checkmark & \checkmark &   &  \\ \hline
noteheadBlack                             & 247741 & \checkmark &   &   &   &   & \checkmark &   & \checkmark & \checkmark & \checkmark & \checkmark & \checkmark & \checkmark &   &  \\ \hline
noteheadDiamondWhole                      & 8      & \checkmark &   &   &   &   & \checkmark &   &   & \checkmark & \checkmark & \checkmark & \checkmark & \checkmark &   &  \\ \hline
noteheadDoubleWholeSquare                 & 3      & \checkmark &   &   &   &   & \checkmark &   &   & \checkmark & \checkmark & \checkmark & \checkmark & \checkmark &   &  \\ \hline
noteheadHalf                              & 8648   & \checkmark &   &   &   &   & \checkmark &   &   & \checkmark & \checkmark & \checkmark & \checkmark & \checkmark &   &  \\ \hline
noteheadTriangleUpBlack                   & 27     & \checkmark &   &   &   &   & \checkmark &   &   & \checkmark & \checkmark & \checkmark & \checkmark & \checkmark &   &  \\ \hline
noteheadTriangleUpHalf                    & 3      & \checkmark &   &   &   &   & \checkmark &   &   & \checkmark & \checkmark & \checkmark & \checkmark & \checkmark &   &  \\ \hline
noteheadWhole                             & 1502   & \checkmark &   &   &   &   & \checkmark &   &   & \checkmark & \checkmark & \checkmark & \checkmark & \checkmark &   &  \\ \hline
noteheadXBlack                            & 81     & \checkmark &   &   &   &   & \checkmark &   &   & \checkmark & \checkmark & \checkmark & \checkmark & \checkmark &   &  \\ \hline
ornamentMordent                           & 7      & \checkmark &   &   &   &   & \checkmark &   &   &   &   &   &   &   &   &  \\ \hline
ornamentTrill                             & 58     & \checkmark &   &   &   &   & \checkmark &   &   &   &   &   &   &   &   &  \\ \hline
ornamentTurn                              & 6      & \checkmark &   &   &   &   & \checkmark &   &   &   &   &   &   &   &   &  \\ \hline
rest                                      & 4      &   &   &   &   &   &   &   &   &   &   &   &   &   &   &  \\ \hline
rest16th                                  & 29411  &   &   &   &   &   &   &   &   &   &   &   &   &   &   &  \\ \hline
rest32nd                                  & 10218  &   &   &   &   &   &   &   &   &   &   &   &   &   &   &  \\ \hline
rest64th                                  & 4      &   &   &   &   &   &   &   &   &   &   &   &   &   &   &  \\ \hline
rest8th                                   & 24124  &   &   &   &   &   &   &   &   &   &   &   &   &   &   &  \\ \hline
restHalf                                  & 1376   &   &   &   &   &   &   &   &   &   &   &   &   &   &   &  \\ \hline
restQuarter                               & 13223  &   &   &   &   &   &   &   &   &   &   &   &   &   &   &  \\ \hline
restWhole                                 & 14382  &   &   &   &   &   &   &   &   &   &   &   &   &   &   &  \\ \hline
slur                                      & 13928  & \checkmark &   &   &   &   & \checkmark &   &   &   &   &   &   &   &   &  \\ \hline
stem                                      & 227889 &   &   &   &   &   & \checkmark &   &   &   &   &   &   &   &   &  \\ \hline
systemicBarline                           & 2078   &   &   &   &   &   &   &   &   &   &   &   &   &   &   &  \\ \hline
tie                                       & 8626   & \checkmark &   &   &   &   & \checkmark &   &   &   &   &   &   &   &   &  \\ \hline
timeSig1                                  & 2      & \checkmark &   &   &   &   & \checkmark &   &   &   &   &   &   &   & \checkmark &  \\ \hline
timeSig2                                  & 1084   & \checkmark &   &   &   &   & \checkmark &   &   &   &   &   &   &   & \checkmark &  \\ \hline
timeSig3                                  & 1203   & \checkmark &   &   &   &   & \checkmark &   &   &   &   &   &   &   & \checkmark &  \\ \hline
timeSig4                                  & 2674   & \checkmark &   &   &   &   & \checkmark &   &   &   &   &   &   &   & \checkmark &  \\ \hline
timeSig5                                  & 507    & \checkmark &   &   &   &   & \checkmark &   &   &   &   &   &   &   & \checkmark &  \\ \hline
timeSig6                                  & 246    & \checkmark &   &   &   &   & \checkmark &   &   &   &   &   &   &   & \checkmark &  \\ \hline
timeSig7                                  & 125    & \checkmark &   &   &   &   & \checkmark &   &   &   &   &   &   &   & \checkmark &  \\ \hline
timeSig8                                  & 1661   & \checkmark &   &   &   &   & \checkmark &   &   &   &   &   &   &   & \checkmark &  \\ \hline
timeSig9                                  & 79     & \checkmark &   &   &   &   & \checkmark &   &   &   &   &   &   &   & \checkmark &  \\ \hline
timeSigCommon                             & 206    & \checkmark &   &   &   &   & \checkmark &   &   &   &   &   &   &   & \checkmark &  \\ \hline
timeSigCutCommon                          & 113    & \checkmark &   &   &   &   & \checkmark &   &   &   &   &   &   &   & \checkmark &  \\ \hline
timeSignatureComponent                    & 443    & \checkmark &   &   &   &   & \checkmark &   &   &   &   &   &   &   & \checkmark &  \\ \hline
tupletBracket                             & 4970   & \checkmark &   &   &   &   & \checkmark &   &   &   &   &   &   &   &   &  \\ \hline
tupletText                                & 4900   & \checkmark &   &   &   &   & \checkmark &   &   &   &   &   &   &   &   & \checkmark   \\ \hline
unpitchedPercussionClef1                  & 223    & \checkmark & \checkmark & \checkmark & \checkmark & \checkmark & \checkmark &   &   &   &   &   &   &   &   &  \\ \hline
wiggleTrill                               & 45     & \checkmark &   &   &   &   & \checkmark &   &   &   &   &   &   &   &   &  \\
\hline
\end{tabular}
\end{table*}

\twocolumn
\section*{Acknowledgement}
The authors acknowledge the support of the AI and Music CDT, funded by UKRI and EPSRC under grant agreement no. EP/S022694/1 and our industry partner Steinberg Media Technologies GmbH for their continuous support especially during the dataset creation.

\end{document}